\documentclass[reviewcopy]{elsart}
\usepackage{amsmath,amssymb}
\usepackage[dvips]{graphicx}
\newcommand{\totder}[2]{\frac{\mathrm{d}#1}{\mathrm{d}#2}}
\newcommand{\ie}{i.e.}
\newcommand{\eg}{e.g.}
\journal{Surface Science}
\begin{document}
\begin{frontmatter}
\title{Diffusion and submonolayer island growth during hyperthermal
deposition on Cu(100) and Cu(111)}
\author[tkk,hu]{M.O.\ Jahma},
\author[tkk]{M.\ Rusanen\corauthref{cor}},
\ead{Marko.Rusanen@tkk.fi}
\author[ks]{A.\ Karim},
\author[hu]{I.T.\ Koponen},
\author[tkk,ri]{T.\ Ala-Nissila}, and
\author[ks]{T.S.\ Rahman}
\corauth[cor]{Corresponding author.}

\address[tkk]{Laboratory of Physics,  P.O.\ Box 1100, Helsinki
University of Technology, FI-02015 TKK, Espoo, Finland}
\address[hu]{Department of Physical Sciences, P.O.\ Box 64,
FI-00014 University of Helsinki, Helsinki, Finland}
\address[ri]{Department of Physics, Box 1843, Brown University,
Providence, RI 02912--1843}
\address[ks]{Department of Physics, Cardwell Hall, Kansas State
University, Manhattan, KS}

\begin{abstract}

We consider the influence of realistic island diffusion rates to
homoepitaxial growth on metallic surfaces using a recently developed rate
equation model which describes growth in the submonolayer regime with
hyperthermal deposition. To this end, we incorporate realistic size and
temperature-dependent island diffusion coefficients for the case of
homoepitaxial growth on Cu(100) and Cu(111) surfaces. We demonstrate that
the generic features of growth remain unaffected by the details of island
diffusion, thus validating the generic scenario of high density of small
islands found experimentally and theoretically for large detachment
rates. However, the details of the morphological transition and scaling
of the mean island size are strongly influenced by the size dependence of
island diffusion. This is reflected in the scaling exponent of the mean
island size, which depends on both temperature and the surface geometry.
\end{abstract}

\begin{keyword}
submonolayer island growth \sep hyperthermal deposition \sep
diffusion \sep copper 

\end{keyword}

\end{frontmatter}
\section{Introduction}

Hyperthermal deposition (HTD) techniques, such as ion-beam assisted
deposition (IBAD) and low-energy ion deposition (LEID) \cite{deg02}
have recently been shown to have great potential in controlling and
improving the properties of thin films as grown by molecular beam
epitaxy. In HTD the island density is larger, the average island size is
smaller \cite{kal97,esc96}, and island size distributions are
considerably broader \cite{deg02,esc96} than in ordinary thermal
deposition. Possible atomistic processes responsible for these effects
include ion enhanced mobilities, cluster dissociation \cite{deg02}, and
defect creation during deposition \cite{kal97,pom02}.

A particularly striking observation made in the LEID experiments is that
with different deposition energies an anomalously high density of small
islands is observed, with the scaled distribution function behaving as
$f(x) \sim 1/x$ for $x < 1$ \cite{deg02}. We have recently shown using
the rate equation (RE) approach \cite{kop04,nev05} that this anomalously
high density of small islands is due to a unique balance between
island--island aggregation and enhanced adatom detachment. These studies
were aimed at describing the generic features of HTD, and thus relatively
idealized approximations for the terms in the rate equations were used to
obtain analytical estimates for the relevant growth exponents and the
scaling function. One of the most important questions that still remains
open is the role of {\em island diffusion}, since for mobile islands the
aggregation rates in the RE approach depend explicitly on the diffusion
coefficients $D_s$ of islands of different sizes $s$. Island diffusion
on surfaces has been studied both theoretically \cite{bog98,hei99,pal99}
and experimentally \cite{wan90,wen94,mor95,pai97,kyu00,cox05,mor05}. While in
the large island limit the size dependence of the island diffusion
coefficients can be classified based on simple basic
processes \cite{kha95}, for smaller islands $D_s$ depends on the
geometric and energetic details of the underlying surface, and can be a
complicated, non-monotonic function of $s$ \cite{san99,tru00}.

In this work, our aim is to explore in detail the influence of realistic
island diffusion coefficients to submonolayer growth with HTD. To this
end, we employ the RE model of Ref.~\cite{kop04} and replace the usually
assumed idealized power law forms of $D_s$ with realistic, temperature
and size--dependent diffusion coefficients $D_s(T)$ for Cu on Cu(100) and
Cu(111) surfaces. These two systems highlight the large differences which
occur for surfaces with different geometry and energetics. We demonstrate
that while the scaling function of the size distribution is largely
unaffected by the details of the diffusion coefficients, the quantitative
values of the growth exponents are sensitive to island diffusion. These
predictions can be easily tested by HTD experiments on different Cu
surfaces.

\section{Model}

In HTD the reversibility of growth is manifested through enhanced adatom
detachment from islands \cite{deg02,pom02}. Detachment and island
mobility allow us to neglect spatial correlations between
islands \cite{kra97,mea88,kan84,kop98}, which justifies the rate equation
description of the problem. Thus, growth is driven by the interplay
between aggregation and detachment, and can be schematically expressed to
be composed of reversible events $A_{i} + A_{j} \rightarrow A_{i+j}$;
$A_{j} \rightarrow A_{1} + A_{j-1}$ between islands of sizes $i$ and $j$
with the rates of aggregation and detachment specified by reaction rates
$K(i,j)$ and $F(i,j)$, respectively. The corresponding REs for the areal
density $n_{s}$ of islands of size $s \geq 1$ read as \cite{kop98,rus03}
\begin{multline}
\totder{n_{s}}{t} =
\frac{1}{2} \sum_{i+j=s}[K(i,j)n_{i}n_{j}-F(i,j)n_{s}]  \\
- \sum_{j=1}^{\infty }[K(s,j)n_{s}n_{j} - F(s,j)n_{s+j}] 
+ \Phi \delta_{1s},
\label{eq:ns}
 \end{multline}
where $\Phi$ is the deposition flux of adatoms in monolayers per second
(ML/s).

The aggregation rate $K(i,j)$ for islands of sizes $i$ and $j$ with
diffusivities $D_{i}$ and $D_{j}$ is given by the Smoluchowski formula
$K(i,j) \propto (D_{i}+D_{j})$, which is also consistent with the point
island approximation used here~\cite{kra97,rus03}. Previously \cite{kop04}
we used a
power-law form for the diffusion coefficients $D_i \sim i^{-\mu}$ with
$1 \le \mu \le 2$ appropriate for island diffusion on metal surfaces. In
order to study the effects of details of the aggregation rates in
observable quantities we  replace the idealized power-law form by
realistic size and temperature dependent diffusion coefficients, as
discussed below in more detail.

The detachment rate of adatoms from islands of size $i+j=s$ depends on
the island size, but only detachment of single adatoms is allowed. Thus
the detachment rate is given by
$F(i,j)=F_{0} (i+j)^{\alpha}(\delta_{1i}+\delta_{1j})$, where the
exponent $\alpha$ is in principle a parameter, but is chosen to be
$\alpha=1/2$ in the present study. In LEID this form for the detachment
rate is physically a well justified choice, because in the regime of
bombarding energies from 10 eV to 100 eV adatom detachment dominates and
island breakup into larger pieces is not expected to
occur \cite{pom02,fra03}. Moreover, since every deposition event at the
vicinity of an island boundary can be assumed to detach adatoms at least
with a probability proportional to the island perimeter (\ie\ $s^{1/2}$),
$\alpha=1/2$ is a reasonable lower limit. This has been also confirmed by
recent Molecular Dynamics simulations on ion bombardment enhanced
detachment in island size region up to 25 atoms where values of
$0.4< \alpha<0.6$ were found \cite{fra03}.

We solve the rate equations using the particle coalescence method
(PCM) \cite{kan84,kra97,rus03}. PCM employs a point-island approximation,
which is valid at low coverages or large island separations. In PCM
aggregation and detachment events occur with probabilities specified by
the corresponding reaction kernels, and the deposition with the rate
proportional to the given adatom flux. An event is then randomly chosen
with a probability weighted by the corresponding rate. Since REs describe
the system in the mean-field limit, there is no information on spatial
correlations in the system. Therefore, in the simulations it is
sufficient to construct a list of islands, which does not correspond to
a physical lattice. To conduct an aggregation event, for example, two
islands are randomly chosen from the list, and an attempt to aggregate
them is made. The complete mixing of the islands required by the
mean-field approximation \cite{kan84} is thus implemented much faster
than including island jumps into empty lattice sites \cite{rus03}.

\section{Diffusion Coefficients}

In reality the diffusion coefficients of islands depend on island size
non-trivially, and not simply as power laws. They can even oscillate as
a function of the island size \cite{wan90,kyu00,hei99}. These
oscillations can be in part understood by the surface geometry: certain
close packed island configurations can be more stable than others on
certain surfaces \cite{tru00}. Here we concentrate on two qualitatively
and quantitatively different types of diffusion coefficients, namely
those of Cu adatom islands on Cu(100) and Cu(111) surfaces.

\begin{figure}
\centering
\includegraphics[angle=-90]{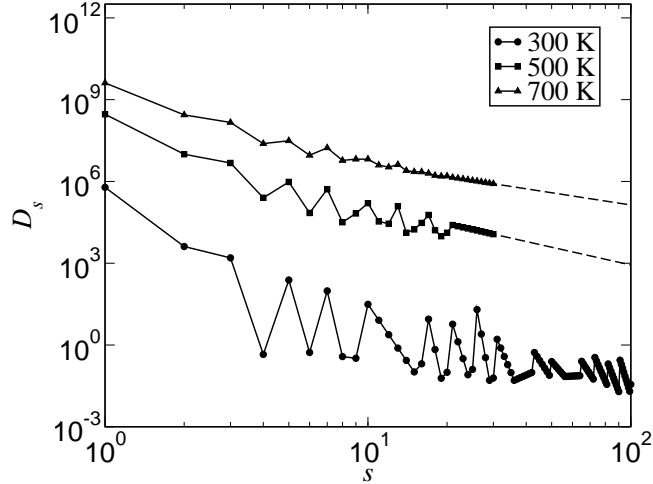}
\caption{Size and temperature dependence of diffusion coefficients 
for 2D adatom islands on Cu(100) at three different temperatures from the
KMC simulations of Ref. \cite{hei99}. For the largest island sizes the
data approximately follow the power law $D_s \sim s^{3/2}$.}
\label{fig:cu100}
\end{figure}

First, the diffusion coefficients $D_s(T)$ from kinetic Monte Carlo (KMC)
simulations, which were based on effective medium theory energetics for
the Cu(100) surface, are shown in Fig.~\ref{fig:cu100} at three different
temperatures $T=300$, $500$, and $700$ K \cite{hei99}. At the lowest
temperatures the oscillations as a function of $s$ are clearly seen. As
temperature increases, the oscillations dampen, and at the highest
temperature the diffusion coefficient is rather well described by 
$D(s) \sim s^{3/2}$, if island diffusion is limited by atomic motion along
the perimeter of an island~\cite{kha95}. For Cu(100), there is a strong
dependence of $D_s$ on temparature here because the single adatom
diffusion barrier is about $0.4$ eV. 

\begin{figure}
\centering
\includegraphics[angle=-90]{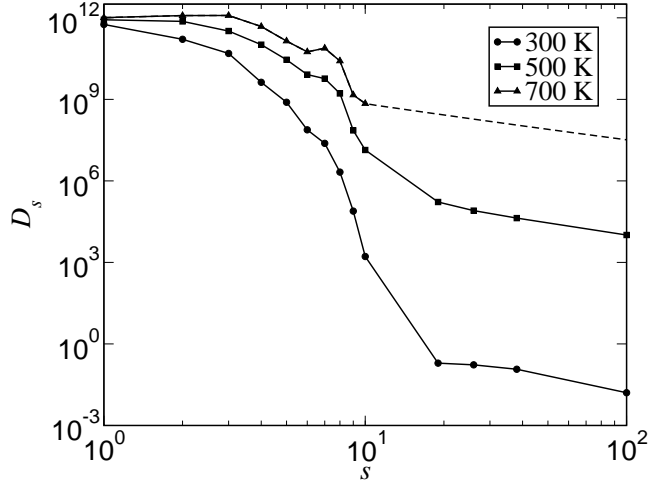}
\caption{Size and temperature dependence of diffusion coefficients 
for 2D adatom islands on Cu(111) at three different temperatures
from the KMC simulations. The dashed lines indicate fit to a $s^{-3/2}$
power law \cite{kha95}.}
\label{fig:cu111}
\end{figure}

In contrast to the behavior on Cu(100), island diffusion on Cu(111)
follows a completely different trend. The most spectacular difference is
the almost complete absence of oscillations in the diffusion coefficient
with size for the smaller islands ($1-10$ atoms) \cite{kar05}. This is
seen to occur when small islands diffuse via occupancy of both hcp and
fcc sites. Also, there is relatively little temperature dependence on
$D_s(T)$ for small sizes, since the surface is relatively smooth with an
adatom activation energy of $0.026$ eV, and almost the same for dimers and
trimers. In the case of Cu(111), the
calculations were carried out using a KMC procedure in which the system
is allowed to evolve according to diffusion processes of its choice. This
is facilitated through the automatic generation of a database of possible
diffusion processes and their activation energy barriers calculated using
embedded atom method potentials. The storage and retrieval of information
from the database is done via a pattern recognition scheme \cite{tru05}.
From simulations performed at $T=300$ K, $500$ K, and $700$ K, the
effective diffusion barriers are found to increase almost monotonically
with size, while the diffusion coefficient takes the form shown in
Fig.~\ref{fig:cu111}. Calculations performed for larger sized islands
($19-100$) show a power law scaling of $D$ with island size and with an
exponent of about $1.57$ which is mildly temperature
dependent \cite{rah04}. This is consistent with the theoretical value of
$3/2$ for large island motion dominated by atom diffusion along island
perimeter \cite{kha95}.

%

\section{Scaling}

In our previous study \cite{kop04} the island size distributions were of
the scaling form and the mean island size had a power-law form.
Therefore, even if the detachment rate was not a homogeneous function of
the island size, we could extract well-defined effective scaling
exponents for the mean island size and the size distribution function.
In the present case, we expect that scaling is invalidated due to
realistic diffusion coefficients which have a naturally inhomogeneous
form. There is no guarantee of either the existence of usual scaling type
of solutions or uniquely defined scaling exponents for the mean
island size and island density. Thus, \eg\ the dynamic exponent for the
mean island size can be expected to be model-dependent, \ie\ we could
obtain different values for different sets of diffusion coefficients.
This has been shown to be the case in more general reversible growth
models \cite{rat94} and growth with mobile islands \cite{kra97,mul01},
where the scaling exponents explicitly depend on the details of the
model.

We define the mean island size $\bar{s} = M_2/M_1$, where the
$k^{\mathrm th}$ moment of the distribution is defined as
$M_k = \sum_{s \ge 1} s^k n_s$. If the scaling of the island size
distribution holds, we expect that
$n_s = \theta \bar{s}^{-2} f(s/\bar{s})$, where
$\bar{s} \sim \theta^{\beta}$, $\theta$ is the surface coverage, and
$f(x)$ is a scaling function independent of coverage.
In our previous work \cite{kop04} we found that the scaling function
is always singular, \ie\ $f(x) \sim 1/x$ for $x < 1$. In order to compare
the scaling function with our previous results, we will use in the
following the modified scaling function of the form $g(x) = x f(x)$.

\section{Results}

The PCM simulations were carried out using island diffusion coefficients
discussed above for Cu(100) and Cu(111), and with the detachment rate
characterized by the exponent $\alpha =1/2$. We define
$\kappa = F_0/D_1$ and $\mathcal{R} = D_1/\Phi$, where $D_1$
is the adatom diffusion coefficient. Parameters $\kappa $ and $R$
denote the importance of detachment relative to diffusion, and of
diffusion relative to deposition flux, respectively. In the simulations
the corresponding values were in the ranges
$10^{-6} \leq \kappa \leq 10^{-1}$ and $10^{5} \leq R \leq 10^{9}$. The
simulations show that for large detachment rates, $\bar{s}$ and $N$
indeed follow a power-law type of behavior, but as $\kappa$ decreases,
$\beta$ becomes coverage-dependent, and saturates only for
$\kappa \rightarrow 0$ as we have previously shown \cite{kop04}. Thus,
only for large $\kappa$ can well-defined scaling exponents be extracted
and regular island growth observed, and in this regime the island size
distributions are of scaling form.

The measured values of the dynamic growth exponent $\beta$ at different
temperatures and large $\kappa$ both for Cu(100) and Cu(111) are shown in
Table~\ref{tab:beta}. The exponent $\beta$ seems to be temperature
dependent for Cu(100) but not for Cu(111). For Cu(100) this can be
explained by examining the corresponding curves for the diffusion
coefficients in Fig.~\ref{fig:cu100}. For small sizes (which is the case
for large $\kappa$) the average slope of the diffusion coefficient curves
depends strongly on temperature. If one fits a power law into the initial
part of the data (for $s \leq 20$), the effective exponents are
$\mu_{\mathrm{eff}} \approx 5.4$, $3.6$, and $2.8$ at $T=300$, $500$, and
$700$ K, respectively. For the power-law type aggregation kernels we
found \cite{kop04} that the dynamic exponent behaves as
$\beta=1/(\alpha+\mu)$, when $K(i,j) \propto i^{-\mu} + j^{-\mu}$. Using
the effective exponents above this prediction gives $\beta \approx 0.17$,
$0.24$, and $0.30$, showing a similar trend as the measured values. If
the measured values are used, somewhat smaller $\mu_{\mathrm{eff}}$ are
obtained than from the fits to the diffusion coefficient data. If the
fitting is done only through maxima of the diffusion coefficient curves,
a better agreement is obtained. It is also interesting to note that
$\bar{s}$ does not depend on $\mathcal{R}$ in this regime, but only on
$\kappa$. This suggests that it could be possible to tune the regime for
the effective diffusion exponent on Cu(100) by changing $\kappa$.
However, this effect is probably too small to be seen, since \eg\ on
Cu(100) at $T=300$ K we get $\bar{s} \approx 12$ atoms for
$\kappa=10^{-3}$, and $\bar{s} \approx 7$ atoms for $\kappa = 10^{-2}$,
while the differences in $\beta$ between these two cases are within the
errors. For Cu(111) there is no clear power-law for small sizes where the
simulation data for diffusion coefficients exist. Instead, small island
sizes seem to be rather mobile relative to adatoms in all temperatures.

\begin{table}
\centering
\begin{tabular}{ccll}
\hline
\hline
$T$ (K) & 300  & 500 & 700 \\
\hline
Cu(100) & 0.25 & 0.34 & 0.43 \\
Cu(111) & 0.25 & 0.22 & 0.29*  \\
\hline
\end{tabular}
\caption{Dynamic scaling exponents $\beta$ as defined in the text at
different temperatures obtained from the PCM simulations, using the
diffusion coefficients for Cu(100) and Cu(111) in Figs.~\ref{fig:cu100}
and \ref{fig:cu111}. The asterisk for Cu(111) at $T=700$ K indicates that
the exponent is not yet constant in time. The errors in other cases are
about $\pm 0.05$.}
\label{tab:beta}
\end{table}
\begin{figure}
\centering
\includegraphics[angle=-90]{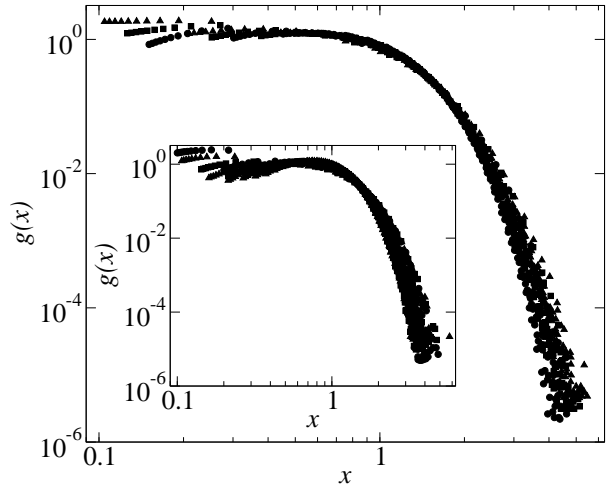}
\caption{The scaled island size distributions on Cu(100). The different
symbols correspond to $T=300$ K (circles), $T=500$ K (squares), and
$T=700$ K (triangles) with $\mathcal{R}=10^6$, $\kappa=10^{-2}$, and
$\theta \le 0.25$ ML. The inset shows the distributions on Cu(111) with
the same parameters.}
\label{fig:ns}
\end{figure}

We expect to observe scaling of island size distributions on the basis of
the fact that well-defined scaling exponents exist for large detachment
rates. In Fig.~\ref{fig:ns} we show scaled island size distributions for
large values of detachment rates using the diffusion coefficients for
Cu(100) and Cu(111) shown in Figs.~\ref{fig:cu100} and \ref{fig:cu111},
and setting $\alpha=1/2$, $\mathcal{R}=10^6$, and $\kappa=10^{-2}$. On
Cu(100) the data for different temperatures collapse to a single curve,
and for $x<1$ the scaled distribution is almost flat. Deviations between
different temperatures occur at small sizes, which reflects the
differences in the diffusion coefficients. The inset shows the scaled
distribution on Cu(111). In this case scaling of the distribution is not
as good, and differences at small sizes are larger than on Cu(100). This
reflects the fact that on Cu(111) the diffusion coefficients for small
island sizes do not follow any power-law type behavior but are
almost constant in $s$, thus invalidating the scaling behavior as in
a more simplified model for $(\mu,\alpha)=(0,1/2)$ \cite{nev05}.
It is, however, expected that deviations could be seen in the large size
tail of the distribution. Since the diffusion coefficients decrease
several orders in magnitude, \eg\ from single atoms to islands of size
$10$, all aggregation events leading to large sizes are basically between
a small island and a large one.

\section{Discussion and Conclusions}

To summarize, we have studied a rate equation model with aggregation and
enhanced adatom detachment corresponding to hyperthermal deposition
conditions using realistic size and temperature dependent diffusion
coefficients for Cu(100) and Cu(111). We have shown that qualitatively
the generic features of growth, \eg\ the form of the scaling function of
the size distribution, are not influenced by diffusion, but quantitative
predictions depend on the microscopic parameters. For example, the mean
island size follows a power-law form in all cases with the exponent
dependent on temperature and surface geometry. Scaling of the size
distribution and the mean island size can be observed, however, only for
large values of adatom detachment rates. These findings suggest that
differences between various surface geometries can be observed only in
the dynamic quantities such as the scaling exponent of the mean island
size, but not \eg\ in the scaled island size distribution which is
time-independent. These predictions can be easily tested with HTD
on Cu surfaces.

{\em Acknowledgments:} This work has been supported in part by the
Academy of Finland through grant 73642 (I.T.K.\ and M.O.J.) 
and through its Center of Excellence program.  The work of T.S.R.\ and
A.K.\ was supported by the National Science Foundation, USA, under grant
ITR-0424677.


\end{document}